\documentclass[12pt,a4paper]{article}

\newcommand{\half}{\frac{1}{2}}

\newcommand{\beq}{\begin{equation}}
\newcommand{\eeq}{\end{equation}}
\newcommand{\bea}{\begin{eqnarray}}
\newcommand{\eea}{\end{eqnarray}}

\newcommand{\dd}{\partial}

\newcommand{\av}[1]{\left\langle #1 \right\rangle}

\title{Taylor's dissipation surrogate and its associated anomaly} 
\author{David McComb\\
School of Physics and Astronomy,\\
University of Edinburgh,\\
EDINBURGH EH9 3JZ.\\
Email: wdm@ph.ed.ac.uk}

\begin{document}
\maketitle

\begin{abstract}
It is shown that, for stationary isotropic turbulence, Taylor's well
known dissipation surrogate $D u'^3/L$ can be derived directly from the
Karman-Howarth equation and is in fact a surrogate for inertial
transfer, which becomes equal to the dissipation, as the Reynolds number
tends to infinity. The expression found for the dissipation rate
$\varepsilon$ is
\[
\varepsilon = \frac{A_3 u'^3}{L}\left[1 + \frac{1}{R_L}\frac{A_2}{A_3}\right],
\]
where the coefficients $A_2$ and $A_3$ depend on the second- and
third-order structure functions respectively and $R_L = u'L/\nu$ is the
Reynolds number based on the integral length scale $L$. Further,
consideration of the spectral energy transfer processes  shows that the
dissipation rate is entirely determined by the energy injection rate.
The role of the viscosity is merely to determine the way in which the
turbulent system adapts to increasing injection rates, both by
increasing the volume of $k$-space and by changing the shape of the
energy spectrum. The quantity $u'^3/L$ is a measure of the inertial
transfer rate and only becomes equal to the dissipation when the
Reynolds number is large enough to permit scale-invariance. It is noted
that similar conclusions hold for shear flows, such as Poiseueille flow,
both laminar and turbulent, where the analogue of energy injection is
the rate of doing work expressed in terms of the pressure gradient and
the mean velocity.
\end{abstract}

\newpage

\section{Introduction}

The idea that the rate of kinetic energy dissipation per unit mass of
fluid turbulence has something anomalous about it, is widespread: see
\cite{Sreenivasan99}\nocite{Falkovich06}\nocite{Cardy08}-\cite{Seoud07} and references
therein. It arises from a result established in 1935 by Taylor
\cite{Taylor35}, who showed, essentially by dimensional arguments, that
the dissipation rate could be written as proportional to $u'^3/l$ where
$u'$ is the rms velocity of the fluid and $l$ is some length scale of
the system. Later, Batchelor \cite{Batchelor71} discussed this idea
further in terms of the decay of kinetic energy (of turbulent motion)
and offered two interpretations of it. The first of these was to see it
as the decay of an amount of energy $u'^2$ in a time $l/u'$. His second
interpretation was to regard it as the effect of an eddy viscosity $ul$
acting on a shear of order $u/l$ to `produce a ``dissipation''
(\emph{sic}) of energy from the energy-containing eddies to smaller
eddies'. It seems likely from the context, and his use of quotation marks,
that Batchelor saw this kind of expression as an approximation to
inertial transfer which would be equal to the \emph{actual} dissipation,
under conditions of local statistical equilibrium. He also made use of
the integral length scale, while noting that it was not as directly
representative of the energy containing eddies as it might be. 

Essentially this is a pragmatic choice and in recent years the integral
length scale $L$ has been generally employed. Taylor's expression is
normally written as
\beq
\varepsilon = D u'^3/L,
\label{taylordiss}
\eeq
where $D$ tends to a constant\footnote{Some workers in the field 
use $C_{\varepsilon}$ for this constant.} with increasing Reynolds number
\cite{Batchelor71}\nocite{Sreenivasan84}\nocite{Sreenivasan98}\nocite{Kaneda03}\nocite{Burattini05}\nocite{Donzis05}-\cite{Bos07},
and is widely used as a surrogate expression for the dissipation.

Taylor's analysis suggested that $D$ would depend on the the geometrical
nature of the boundary conditions; while Batchelor pointed out that it
might depend on the time of decay, the initial conditions of the
turbulence, and the choice made for $l$. This view has received recent
theoretical support \cite{Mazellier08},\cite{Gotoh09}. 
 
Now let us turn our attention to the so-called \emph{dissipation anomaly}.
Given that dissipation is due to viscosity and that the dissipation rate
is formally defined in terms of the coefficient of viscosity, thus:
\beq
\varepsilon = \av{\frac{\nu}{2}\left(\frac{\dd u_i}{\dd x_j} + \frac{\dd
u_j}{\dd x_i}\right)^2},
\label{dissdefn}
\eeq
the perceived anomaly takes one of two related forms:
\begin{enumerate}
\item The fact that the dissipation rate, as given by (\ref{taylordiss})
and verified by experiment, is found to be independent of the fluid
viscosity
\cite{Sreenivasan84}\nocite{Sreenivasan98}\nocite{Kaneda03}\nocite{Burattini05}\nocite{Donzis05}-\cite{Bos07},
providing that the Reynolds number is sufficiently large.
\item The existence of finite dissipation in the limit of vanishingly
small viscosity (or the limit of infinite Reynolds number)
\cite{Sreenivasan99}\nocite{Falkovich06}-\cite{Cardy08}. 
\end{enumerate}

Noting that (\ref{taylordiss}) has only been derived from dimensional
considerations, and analysed qualitatively in terms of the kinetic
energy $u'^2$ and the eddy turnover time $L/u'$, there is a need for a
more theoretical analysis. From our present point of view, the recent
work of Doering and Foias \cite{Doering02} is of interest. For the case
of forced turbulence, they have established both upper and lower bounds
on the dissipation rate. We shall return to this work at appropriate
points in our own analysis.

In this paper we mainly confine our attention to stationary, isotropic
turbulence. We examine the exact relationships expressing conservation
of energy, first in real space, and then in wavenumber space. We show
that Taylor's expression follows quite naturally from the Karman-Howarth
equation, as the Reynolds number increases. In the process, we find that
it is a surrogate for the inertial transfer, and not for the dissipation
rate, as such.  Then, by considering the equivalent $k$-space relation,
we verify that the rate of doing work by body forces is the controlling
quantity which determines the dissipation rate in this kind of
turbulence; and by, fairly obvious extension, in all fluid flows.

\section{Taylor's surrogate and the Karman-Howarth equation} 

The Karman-Howarth equation may be written in terms of structure
functions as \cite{Landau59}:
\beq
-\frac{2}{3}\varepsilon - \half \frac{\dd S_2}{\dd t} = \frac{1}{6
r^4}\frac{\dd}{\dd r}(r^4  S_3) - \frac{\nu}{r^4}\frac{\dd}{\dd
r}\left(r^4 \frac{\dd S_2}{\dd r}\right),
\label{khe}
\eeq
where the structure function of order $n$ is given by
\beq
S_n = \av{\left(u(x+r) -u(x)\right)^n}.
\eeq

In this section we first consider the stationary case; then the
relationship of the work we present here to other recent work on the
subject; and finally the extension of our analysis to freely decaying
turbulence. 

\subsection{Stationary turbulence}

For the case of stationary turbulence, we may set the time-derivative
equal to zero, and re-arrange the Karman-Howarth equation to obtain an
expression for the dissipation, thus:
\beq
\varepsilon = - \frac{1}{4 r^4}\frac{\dd}{\dd r}(r^4  S_3) + \frac{3\nu}{2r^4}\frac{\dd}{\dd
r}\left(r^4 \frac{\dd S_2}{\dd r}\right).
\label{khdiss}
\eeq
Now make the \emph{change of variables},
\beq
S_n = V^n f_n(x), \qquad \mbox{with} \qquad x=\frac{r}{b},
\label{dimless}
\eeq
where $V$ is some constant velocity scale and $b$ is any length scale.
We should note that this step involves no approximations or non-trivial
assumptions. It merely introduces the $f_n$ as dimensionless forms of the 
structure functions. As they are dimensionless, their dependence on $r$
must be scaled by some length, here denoted by $b$. In Section 2.3 we
will discuss the introduction of \emph{self-similar} and
\emph{similarity solutions} when we extend the present analysis to
time-dependent (i.e. freely decaying) turbulence.

With these substitutions, equation (\ref{khdiss}) becomes:
\beq
\varepsilon = \frac{A_3 V^3}{b} + \frac{A_2\nu V^2}{b^2},
\label{interdiss}
\eeq
where the coefficients $A_3$ and $A_2$ are given by
\beq
A_3 = -\frac{1}{4x^4}\frac{\dd}{\dd x}\left[x^4 f_3(x)\right],
\label{a3coefft}
\eeq
and
\beq
A_2 = \frac{3}{2x^4}\frac{\dd}{\dd x}\left[x^4 \frac{\dd f_2}{\dd x}\right].
\eeq
Taking out common factors, it is readily seen that the expression for
the dissipation becomes
\beq
\varepsilon = \frac{A_3 V^3}{b}\left[1 +
\frac{1}{R_b}\frac{A_2}{A_3}\right],
\label{vbdiss}
\eeq
where the Reynolds number is given by $R_b = V b / \nu$. We note that
this equation is still just the Karman-Howarth equation: no
approximation has been made.

Now we may identify the prefactor on the right hand side of
(\ref{vbdiss}) as being the same as the Taylor dissipation
surrogate (\ref{taylordiss}); providing we make the choices $V = u'$,
the root-mean-square velocity, and $b=L$, the integral scale. Then we
obtain the general relation
\beq
\varepsilon = \frac{A_3 u'^3}{L}\left[1 +
\frac{1}{R_L}\frac{A_2}{A_3}\right],
\label{finaldiss}
\eeq
but now the Reynolds number is given by $R_L = u' L / \nu$. Note that the
$f_n$ are determined by this choice and hence also the coefficients 
$A_3$ and $A_2$.

Clearly, as the Reynolds number goes to infinity, this expression
reduces to Taylor's surrogate for the dissipation (\ref{taylordiss}),
provided that $A_3$ becomes a constant. However, as it represents the
inertial-transfer term, we should really describe it as Taylor's
surrogate for inertial transfer.

\subsection{Comparison with other work}

It is also of interest to compare equation (\ref{interdiss}), which is
an intermediate stage in our calculation, to the result for an upper
bound on the dissipation as given by Doering and Foias \cite{Doering02}.
This is featured in their abstract as
\[
\varepsilon \leq c_1 \nu \frac{u'^2}{l^2} +c_2 \frac{u'^3}{l},
\]
and corresponds to their equation (40). Here the coefficients $c_1$ and
$c_2$ depend on the shape of the forcing function, while $l$ is its
longest length-scale. Obviously this is quite different from our own
result, where the corresponding parameters depend on the fluid
turbulence and not on the forcing, and we have an equality, rather than
an inequality. Nevertheless, the general similarity of the two results
is worthy of notice.

After some further manipulations, these authors introduce a function
$\beta$ (the same as our $D$ in (\ref{taylordiss})) such that
\beq
\beta \equiv \frac{\varepsilon l}{U^3} \leq \left(\frac{a}{Re} + b
\right),
\label{doering}
\eeq
where the symbols are all as in \cite{Doering02}. They then use the identity
$Re = \beta R^2_{\lambda}$ to substitute for $Re$, solve the resulting
quadratic equation, and obtain
\beq
\beta \leq \frac{b}{2} \left[1 + \sqrt{1 + \frac{4a}{b^2 R^2_{\lambda}}} \right],
\eeq
in terms of the Taylor-Reynolds number $R_{\lambda}$.

An interesting development is that Donzis, Sreenivasan and Yeung
\cite{Donzis05} have taken this upper bound as an equality, which they
write as 
\beq
\beta = A \left( 1 + \sqrt{1 + (B/R_{\lambda})^2}\right).
\label{sreeni}
\eeq
They fit this curve to results obtained from a numerical simulation and
obtain an impressively close fit, with $A\sim 0.2$ and $B \sim 92$,
leading to an asymptotic value of $\beta = 0.4$.

Evidently our equation (\ref{finaldiss}), being readily reduced to the
same form as (\ref{doering}), may be further reduced to the same form as
({\ref{sreeni}), with
\beq
A = A_3/2 \qquad \mbox{and} \qquad B = 2 A^{1/2}/A_3.
\eeq
Thus the implication is that it also agrees well with the results from
simulation. We are currently doing our own numerical calculations; but,
in the meantime, it is reassuring to know this.

\subsection{Extension to freely decaying turbulence}

We now consider whether our results can also be applied to freely decaying
isotropic turbulence. The neglect of the time-derivative term in
(\ref{khe}) is usual, even for freely decaying turbulence, provided that
the Reynolds number is large enough and one restricts attention to the
inertial range. For instance, this step is required in order to
derive the well-known `4/5' law and is known as \emph{local
stationarity}.

Let us consider the effect on (\ref{dimless}) of choosing $V$ and $b$ to
be the Kolmogorov velocity and length scales, as given by:
\beq
v = (\nu \varepsilon)^{1/4} \qquad \mbox{and} \qquad \eta =
(\nu^3/\varepsilon)^{1/4}:
\label{kolscales}
\eeq
see equation (21.4) in \cite{Monin75} or equations (2.131), (2.132) in
\cite{McComb90}. For $n=2$, substitute (\ref{kolscales}) into
(\ref{dimless}) for $V$ and $b$, respectively. We obtain
\beq
S_2(r) = \nu^{1/2}\varepsilon^{1/2} f_2(r/\eta).
\eeq
If we are to make any further progress, then we must assume a (specific)
\emph{self-similar} form for $f_2$. That is, we assume:
\beq
f_2(r/\eta) = \frac{1}{\eta^{2/3}}f_2(r),
\eeq
from which the second-order structure function becomes:
\beq
S_2(r) = \varepsilon^{2/3}f_2(r); \qquad f_2(r) = r^{2/3}.
\eeq
Similarly, we can show that the third-order structure function is of the
form 
\beq
S_3(r) = \varepsilon f_3(r); \qquad f_3(r) = r.
\eeq
It should be noted that the forms in (\ref{kolscales}) omit constants of
order unity and this is reflected in these results. Also, we do not
consider the question of non-canonical exponents or intermittency
corrections as these matters have been addressed elsewhere \cite{McComb09a}.

Evidently our analysis is consistent with the K41 picture and should
apply for the case of local stationarity. Let us now consider a more
general situation, where we assume that the decaying turbulence is 
self-preserving. Now we replace (\ref{dimless}) by the time-dependent
form: 
\beq
S_n(r,t) = u'^n(t_e) f_n(x),
\label{tdimless}
\eeq
where
\beq
x=r/L(t).
\eeq
Note that our choice of an initial time for a \emph{similarity solution}
requires some care. We have taken it to be some $t=t_e$, when the
turbulence is said to have evolved from arbitrary initial conditions. If
we assume that the turbulence is self-preserving after $t=t_e$, then
there is no explicit time-dependence in the dimensionless structure
functions, and equation (\ref{finaldiss}) applies in the present case as
well.

In practice self-preservation is only likely to be approximately
correct, and we can take account of the residual time-dependence as a
perturbation of the similarity solution, writing  equation
(\ref{tdimless}) as
\beq
S_n(r,t) = u'^n(t_e) f_n(x,\tau),
\eeq
where now
\beq
\tau = t/T; \qquad T=L(t)/u'(t_e).
\eeq
In these circumstances it is easily shown that equation
(\ref{finaldiss}) can be generalised to the form:
\beq
\varepsilon = \frac{(A_3 - B_2)u'^3}{L}\left[1 +
\frac{1}{R_L}\frac{A_2}{A_3 - B_2}\right],
\label{finaldisstime}
\eeq
where the new coefficient $B_2$ is given by
\beq
B_2 = \frac{3}{4}\frac{\dd f_2}{\dd \tau},
\eeq
for small values of the time derivative.

In all, this suggests that equation (\ref{finaldiss}), although derived
for stationary turbulence, may apply quite well to decaying turbulence.

\section{The spectral picture}

In order to examine these ideas further,  we study the spectral
energy balance in wavenumber space, and consider the effect of
increasing the Reynolds number, ultimately taking the limit of infinite
Reynolds numbers. The three energy transport processes, viz., injection,
dissipation and inertial transfer will now be considered in turn.

We begin with the energy balance equation, in its well known form,
\begin{equation}
	\left( \frac{\partial}{\partial t} + 2\nu k^2 \right) E(k,t)
	= T(k,t) + W(k),
\label{enbalt}
\end{equation}
where the energy transfer spectrum $T(k,t)$ is given by:
\begin{equation}
	T(k,t) = 2 \pi k^2  M_{\alpha\beta\gamma} (\mathbf{k}) \int
	\mbox{d}^3j \,  \left\{ C_{\beta\gamma\alpha}
	(\mathbf{j},  \mathbf{k}-\mathbf{j}, -\mathbf{k},t) - C_{\beta\gamma\alpha}
	(-\mathbf{j},  -\mathbf{k}+\mathbf{j},\mathbf{k},t) \right\},
\label{tseqn}
\end{equation}
and $C_{\alpha\beta\gamma}(\mathbf{k,j,\mathbf{k}-\mathbf{j}})$ is the three-velocity
correlation. We may write $T(k,t)$ as
\beq
T(k,t) =  \int^{\infty}_{0}S(k,j;t)\,dj,
\eeq
where $S$ depends on the triple moment: its form can be deduced from (\ref{tseqn}).
It can be shown that $S$ is antisymmetric under the interchange
$k\rightleftharpoons j$:
\beq
S(k,j;t) = -S(j,k;t).
\eeq
Hence
\beq
\int^{\infty}_{0}T(k,t)dk = \int^{\infty}_{0}dk\int^{\infty}_{0}dj
\,S(k,j;t) = 0,
\label{encon}
\eeq
is an exact symmetry which expresses conservation of energy.

In order to study the stationary case, we have added an input spectrum
$W(k)$ (if one wishes, this can be related to the covariance of the
random stirring forces \cite{McComb90}). We also introduce
$\varepsilon_W$ as the rate at which the stirring forces do work on the
turbulent fluid:
\beq
\varepsilon_W =\int^{\infty}_{0}W(k)dk.
\eeq

The dissipation rate $\varepsilon_D$ in $k$-space is defined by 
$\varepsilon_D = -dE/dt$ for freely decaying turbulence. As is well
known, we can obtain an expression which is also valid for the
stationary case, in the usual way, by temporarily setting $W(k)=0$,
integrating (\ref{enbalt}) over wavenumber, and rearranging, such that
the energy balance becomes:
\beq
\varepsilon _D = - \frac{dE}{dt} =  \int^{\infty}_{0}2\nu k^{2}E(k,t)dk,
\eeq
where we have also invoked equation (\ref{encon}). The region in $k$-space
where the dissipation mainly occurs is characterised by the Kolmogorov
dissipation wavenumber:
\beq
k_D=(\varepsilon_D/\nu^3)^{1/4}.
\label{kdisswave}
\eeq

Then, restoring the injection spectrum, for the stationary case we
have $dE(k,t)/dt = 0$, and the energy balance becomes:
\beq
T(k) + W(k) - 2\nu k^{2}E(k) = 0.
\label{statbal}
\eeq
Integrating both sides with respect to wavenumber, we have:
\beq
\int^{\infty}_{0}W(k)dk - \int^{\infty}_{0}2\nu k^{2}E(k)dk =0; \quad
\mbox{or:} \quad
\varepsilon_W = \varepsilon_D,
\label{workbal}
\eeq
as a rigorous consequence of stationarity. In other words, for a
stationary flow, the dissipation is governed by the rate at which we do
work on the fluid in order to produce a required fluid motion. We shall
enlarge on this point presently.

\subsection{Inertial transfer and scale-invariance}

We now consider inertial transfer of energy in $k$-space.  The energy
flux is introduced when we integrate each term of (\ref{enbalt}) with
respect to wavenumber, from zero up to some wavenumber $\kappa$.
Reverting for the moment to the general (i.e. non-stationary) case,  we
obtain:
\beq
\frac{d}{dt}\int_{0}^{\kappa} dk\, E(k,t)  = 
- \int^{\infty}_{\kappa} dk\,\int^{\kappa}_{0} dj\, S(k,j;t)
-2 \nu\int_{0}^{\kappa} dk\, k^2 E(k,t),
\label{fluxbalt}
\eeq
where we have used the antisymmetry of $S$, and made some
rearrangements. In this form the effect of the transfer term is readily
interpreted as the net flux of energy from wavenumbers less than
$\kappa$ to those greater than $\kappa$, at any time $t$
\cite{Batchelor71}. Denoting this flux by  $\Pi(\kappa)$; and, in order
to avoid the ambiguity associated with the scale-invariance paradox, 
making an exact decomposition of the transfer spectrum into
filtered-partitioned forms $T^{+-}(k|\kappa)$ and $T^{-+}(k|\kappa)$
\cite{McComb08}, we have
\beq
\Pi (\kappa) = \int^{\infty}_{\kappa} dk\, T^{+-}(k|\kappa) =-\int^{\kappa}_{0} dk\,
T^{-+}(k|\kappa),
\label{tp}
\eeq
where we have now assumed stationarity and dropped the time dependence.
(Note that the decomposition is completed by $T^{--}(k|\kappa)$ and
$T^{++}(k|\kappa)$, which are separately conservative on the intervals
$[0,\kappa]$ and $[\kappa,\infty]$, respectively \cite{McComb08}.)
The maximum value of the energy flux is $\Pi_{max}(\kappa)$, where
$T^{+-}(k|\kappa) = T^{-+}(k|\kappa) =0$.

When setting up stationary, isotropic turbulence by means of an
arbitrary choice of stirring forces, it is usual to try to reproduce the
characteristic features of the classic turbulent shear flows. In terms
of the energy spectrum, these may be seen as the energy-containing range
(low wavenumbers), the inertial range (intermediate wavenumbers) and the
dissipation range (large wavenumbers). It has been known since the late
1930s \cite{Batchelor71} that the energy-containing and dissipation
ranges become more widely separated as the Reynolds number is increased;
with the Kolmogorov dissipation wavenumber $k_D$, as given by 
(\ref{kdisswave}), providing a reliable measure of this process. In
practice, what this means is that we should choose the injection
spectrum $W(k)$ to have a somewhat peaked form at low values of
wavenumber, when compared to $k_D$.

We may formalize this picture as follows. At sufficiently large
$R_{\lambda}$, the energy-containing and dissipation ranges become
separated by the inertial range of wavenumbers, thus:
\beq
k_{bot}\leq \kappa \leq k_{top},
\label{irlim}
\eeq
where $\kappa$ now stands for any wavenumber in the inertial range.
In this case, the injection and dissipation spectra satisfy approximate
relationships as follows:
\beq
\int^{k_{bot}}_{0}dk W(k)\simeq\varepsilon_{W};\quad
\mbox{and}\quad
\int^{\infty}_{k_{top}}dk\, 2\nu k^{2}E(k) \simeq\varepsilon_D.
\eeq
In this range of wavenumbers the maximum energy flux should be
approximately constant and we
will find it helpful to introduce a specific symbol for this quantity,
thus:
\beq
\Pi_{max}= \varepsilon_T.
\eeq
For stationarity, we must have the overall energy balance:
\beq
\varepsilon_W = \varepsilon_T = \varepsilon_D.
\label{three}
\eeq
We note that these three different physical processes are normally
denoted by the single symbol $\varepsilon$, this being justified by
their all being numerically equal. In our view, it is necessary to draw
a distinction between them in order to avoid confusion. 

It should now be apparent that Taylor's expression, as given by
(\ref{taylordiss}), should be written as $\varepsilon_T = D u'^3/L$;
which also explains the observed dependence on Reynolds number (see Fig.
1 in \cite{Sreenivasan98}). In other words, (\ref{taylordiss}) becomes
equal to the dissipation when (\ref{three}) is satisfied.

\subsection{The limit of infinite Reynolds numbers}

The separation of the energy-containing and dissipation ranges by a
scale-invariant inertial range has an interesting consequence. It allows
us to obtain separate low-$k$ and high-$k$ balance equations; by first
integrating from zero up to $k_{bot}$, and then from infinity down to
$k_{top}$. First, we have
\beq
\int^{k_{bot}}_{0}dk\int^{\infty}_{k_{bot}}dj\, S(k,j) +
\int^{k_{bot}}_{0}W(k)dk = 0.
\label{lowbal}
\eeq
That is, energy supplied directly by the input term to modes with
$k\leq k_{bot}$ is transferred by the nonlinearity to modes with
$j\geq k_{bot}$. Thus $T(k)$ behaves like a \emph{dissipation} and absorbs energy. Second,
\beq
\int^{\infty}_{k_{top}}dk\int^{k_{top}}_{0}dj\, S(k,j) -
\int^{\infty}_{k_{top}}2\nu k^{2}E(k)dk = 0.
\label{highbal}
\eeq
That is, the nonlinearity transfers energy from modes with $j\leq k_{top}$ to
modes with $k\geq k_{top}$, where it is dissipated into heat. In this
range of wavenumbers $T(k)$ behaves like a \emph{source} and emits energy
which is then dissipated by viscosity.

Now we reach the crux of our argument. We examine the limit of infinite
Reynolds number, as follows. We keep the injection spectrum $W(k)$ fixed
(and hence also $\varepsilon_W=\varepsilon_T=\varepsilon_D$ are held
constant), and allow the viscosity to tend to zero. As a result
(originally pointed out by Batchelor in 1953 \cite{Batchelor71}), the
sink of energy is displaced to $k=\infty$, in the limit of infinite
Reynolds number. We can deduce that this is so, either from the form of the
Kolmogorov dissipation wavenumber or the form of the dissipation term;
or even from a consideration of the local Reynolds number for mode $k$
\cite{Batchelor71}.

However, we cannot emphasise too strongly that this has nothing to do
with the Euler equation. The Euler equation can indeed be obtained from
the Navier-Stokes equation by setting the viscosity equal to zero. The
result is the equation of motion for an inviscid, zero-dissipation
fluid. Here, in contrast, we take the \emph{limit} of zero viscosity,
while keeping the dissipation constant. 

This operation was later formalized by Edwards \cite{Edwards65} (or see
Section 6.2.7 in \cite{McComb90}), who argued that the dissipation at
infinity could be represented by a Dirac delta function, thus:
\beq
\left. \int^{\infty}_{0}\,dk\,\lim_{\nu \rightarrow 0}\,  2\nu k^2 E(k)\right|_{\varepsilon_D = \mbox{constant}} =
\int^{\infty}_{0}\,dk\,\varepsilon_D\, \delta(k-\infty) =\varepsilon_D.
\eeq
Similarly, with an infinite extent of wavenumber space, any injection
spectrum of finite extent can be scaled back to the origin, so that we
have also:
\beq
\left. \int^{\infty}_{0}\,dk\, \lim_{\nu \rightarrow 0}\, W(k)\right|_{\varepsilon_W = \mbox{constant}} =\int^{\infty}_{0}\,dk\,\varepsilon_W\,
\delta(k) = \varepsilon_W. 
\eeq
Thus, in the limit of infinite Reynolds numbers, equation
(\ref{statbal}) for the spectral energy balance may be written as:
\beq
-T(k) =  \varepsilon_W\,\delta(k) - \varepsilon_D\, \delta(k-\infty).
\eeq
Note that these forms satisfy all the relevant relationships given above
as equations (\ref{irlim}) - (\ref{highbal}) and, although this may seem a rather extreme
procedure, it is in fact nothing more than a different mathematical
representation of scale invariance in the inertial range.

The general forms of our arguments given here in spectral space are not
unlike those of Davidson (see Section 3.2.2 in \cite{Davidson04}), which
are presented for physical or scale space. A particular point of
interest is his example of a sudden expansion in a pipe, where the `head
loss' is determined by continuity and the momentum theorem (plus, it
must be said, an approximation based on the existence of stagnation
points) but not apparently on the fluid viscosity.

Also, it is of interest to note that our views expressed here seem to be
quite close to those of Doering and Foias \cite{Doering02}, despite the
very different form of the analysis. Essentially both treatments give a
dominant role to the production of the turbulence and, for instance, our
equation (\ref{workbal}) is essentially just equation (17) of reference
\cite{Doering02}. 

\section{Conclusion}

From this simple analysis, it is apparent that there are really no
grounds for viewing the lack of dependence of the dissipation on the
viscosity as an anomaly. Conservation of energy ensures that the
dissipation is just equal to the injection rate. The role of the
viscosity is to control the energy occupation of wavenumber space, along
with the shape of the spectrum. Noting that Taylor's expression for the
dissipation, as given by equation (\ref{taylordiss}), is really just the
inertial transfer rate, we can appreciate that it will depend on
Reynolds number, until the latter is large enough for the
energy-containing and dissipation ranges of wavenumber to be adequately
separated, to the point where inertial transfer is well-defined, and
equal to the actual dissipation rate. Also, for the limit of infinite
Reynolds number, it can be seen that the symmetry-breaking effect of the
viscosity is still present in the form of the delta functions, as in
equation (31).

We can also attempt to say something more general about shear flows. If
we consider classical shear flows, then laminar Poiseuille flow
presents one of the few cases where the Navier-Stokes equation may be
solved exactly. This is because the nonlinear term vanishes identically,
and the problem resolves itself into a balance between viscous and
pressure forces. Here the dissipation rate can be worked out exactly,
and is found to be independent of viscosity. It is, as it must be, equal
to the injection rate; which can also be worked out exactly, in terms of
the rate at which the pressure force moves its point of application. In
this case, the gradient of the velocity depends inversely on the
viscosity, and becomes steeper in order to maintain the dissipation rate
as the viscosity is decreased at constant axial pressure gradient. This
is, of course, analogous to the behaviour of the Kolmogorov length scale
under increasing Reynolds number at constant injection rate. 

Extension of the analysis to turbulent Poiseuille flow requires a
little more work, but those nonlinear terms which do not vanish
identically, do vanish when integrated over the system volume; and the
analysis for the laminar case applies here as well.

\section*{Acknowledgements}

It is a pleasure to acknowledge the continuing support provided by the
award of an Emeritus Fellowship by the Leverhulme Trust, along with the 
hospitality of the Isaac Newton Institute (in the form of a Visiting
Fellowship) and that of the Institute for Mathematical Sciences,
Imperial College. I would also like to thank Arjun Berera, Francis
Barnes, Claude Cambon, Stuart Coleman, Gregory Falkovich, Jorgen
Frederiksen, Matthew Salewski, Katepalli R. Sreenivasan, Christos
Vassilicos, and Sam Yoffe, for stimulating discussions, helpful comments
or other assistance.


\end{document}